\documentclass{article}
\usepackage[margin=1in]{geometry} 
\usepackage[T1]{fontenc}
\usepackage{graphicx}
\usepackage{natbib}
\usepackage{amsfonts,amsmath,amssymb}
\usepackage[hidelinks]{hyperref}
\usepackage{xcolor}
\hypersetup{
    colorlinks,
    linkcolor={red!50!black},
    citecolor={blue!50!black},
    urlcolor={blue!80!black}
}

\begin{document}
\title{Beyond Human Intervention: Algorithmic Collusion\\ through Multi-Agent Learning Strategies}
%
%
\author{Suzie Grondin$^1$ \and Arthur Charpentier$^{2}$\thanks{AC acknowledges the National Sciences and Engineering Research Council (NSERC) for funding (RGPIN-2019-07077) and the SCOR Foundation for Science.} \and Philipp Ratz$^{2}$}

\date{
	$^1$ENSAE Institut Polytechnique \\ \texttt{suzie.grondin@ensae.fr}\\%
	$^2$Universit\'e du Qu\'ebec \`a Montr\'eal (UQAM) \\ \texttt{charpentier.arthur@uqam.ca}\\
    \texttt{ratz.philipp@courrier.uqam.ca}\\[2ex]
    }

\maketitle             
\begin{abstract}
Collusion in market pricing is a concept associated with human actions to raise market prices through artificially limited supply. Recently, the idea of \emph{algorithmic} collusion was put forward, where the human action in the pricing process is replaced by automated agents. Although experiments have shown that collusive market equilibria can be reached through such techniques, without the need for human intervention, many of the techniques developed remain susceptible to exploitation by other players, making them difficult to implement in practice. In this article, we explore a situation where an agent has a multi-objective strategy, and not only learns to unilaterally exploit market dynamics originating from other algorithmic agents, but also learns to model the behaviour of other agents directly. Our results show how common critiques about the viability of algorithmic collusion in real-life settings can be overcome through the usage of slightly more complex algorithms. 
\end{abstract}

\section{Introduction}

Reinforcement Learning (RL) has received considerable attention in the research community in recent years. Through the ability to solve complex dynamic problems, it has been used to optimise energy production \cite{grinberg2014optimizing}, helped navigating autonomous aviation \cite{bellemare2020autonomous} and also to improve large language models \cite{lee2023rlaif}. In the field of economics, RL was recently used for policy design \cite{zheng2021ai} or to study the pricing behaviour of automatic agents in \cite{calvano2020artificial}. Specifically, the latter found that algorithms in a simple economic environment can autonomously learn to resort to pricing strategies not unlike what could be observed under a market outcome where producers collude on prices. These results have lead to a discussion among scholars, as it potentially poses problems to existing policy regulating pricing decisions. 

Follow-up studies have extended the understanding of the mechanisms with both experimental and theoretical results. For example, \cite{aryal2022coordinated,clark2023algorithmic} used empirical case studies and added weight to the claims raised in \cite{calvano2020artificial}, both found that the introduction of RL-based pricing methods can indeed raise prices for consumers. However, a drawback of these studies is that they cannot differentiate between pricing models themselves and potentially collusive pricing strategies, as better (static) pricing models might be introduced at the same time as (dynamic) reinforcement learning agents, mixing the results. More theoretical work by \cite{banchio2022adaptive} has focused on the conditions under which supracompetitive prices can arise between agents in a two-player game. However, they stay in much the same setup as \cite{calvano2020artificial}, which makes it difficult to derive more general conclusions. 

This article extends upon this literature, and aims address some of their weaknesses. We find that many results in the standard setup are heavily dependent on the choice of parameters, but that this can be overcome by adapting the objective function to two objectives. In particular, we analyse the situation where the experiments are not symmetric, that is, a setup where parameters differ across agents. In previous studies, experimental parameters were changed in the global experimental setup, but kept equal across agents. We show that this assumption goes beyond simply having access to the same modelling techniques, as it would also require market participants to make the same choices when initialising their agents. Departing from this assumption shows that obtaining an equilibrium where \emph{all} agents make a supracompetitive profit is difficult to obtain, contrary to previous results. In theory, a collusive equilibrium could still be obtained either the usage of the same \emph{pretrained} algorithm (for example, provided by a third party) or having an agreement on the selected hyperparameters. However, a recent brief by the Federal Trade Commission (FTC) \cite{ftc_brief} argues that both of these practices would fall under the scope of existing regulation and can be considered as banned anti-competitive practices. 

An interesting concept, developed by \cite{rocher2023adversarial}, is that of \emph{adversarial collusion}, where a participant in the market unilaterally exploits regularities in the pricing algorithms of competitors to achieve supracompetitive profits for \emph{all} market participants. This poses a direct issue to existing regulation, as it does not require the participants to have an agreement on the choice of algorithms, nor the choice of hyperparameters and it is unlikely to entail a response from other competitors, as profits increase across the market. We expand upon this idea in a more general setting and reformulate the problem as a multi-objective optimisation, which allows more general agents to be developed. After these adjustments, we arrive at much the same conclusions about market equilibria as previous studies, but with added robustness.  

As we will discuss below, both algorithmic collusion and adversarial collusion can be considered under a similar angle through their reward function. The reward function essentially regulates how an agent in an RL problem assigns value to an outcome at a given time, which includes discounting expected future revenues. Whereas in the computer science literature the choice of the reward function, and its usefulness is acknowledged, but also debated \cite{silver2021reward,vamplew2022scalar}, the economic literature currently largely abstracts from this discussion. Here, we aim to bridge this gap. We follow \cite{charpentier2021reinforcement} and introduce price setting as a more formal game that is solved by reinforcement learning agents. This allows us to discuss critiques and conclusions of recent research in a unifying form. In particular, we suggest that findings from previous research on cartel-formation \cite{levenstein2006determines} should be used to re-address competition policy. Whereas in the past, forming cartels or collusive market outcomes was hindered by rapidly changing market conditions and difficulties in high costs to establish such prices in the first place, modern algorithms can quickly adjust and analyse large amounts of data in shorter periods lowering these costs. Further, as RL agents inherently set prices with a long(er)-term objective, a wide spread introduction of RL-based pricing mechanisms might further reinforce the convergence to supra-competitive prices, which should be addressed through suitable policy. 

The remainder of the article is structured as follows: Section \ref{sec:background} introduces the basic concepts of reinforcement learning and multi-agent reinforcement learning with a focus on economic applications. We then start the discussion on the implications using simple games that have been well-studied in the past in Section \ref{subsec:econgames} and extend the discussion to more complex games in Section \ref{sec:main} to show how RL can be used to find solutions that can be detrimental for consumers. Finally, we discuss our findings in Section \ref{sec:conclusion}.

\section{Background on Reinforcement Learning}\label{sec:background}

The field of Reinforcement Learning is too vast to be discussed in its entirety, and we refer to for example \cite{rl_book}, whose book we largely follow here, for a more general overview. For our purposes, we focus on the formalisation of multi-player games in a context that can be resolved using RL-algorithms. In particular and as was done in previous studies, we restrict our analysis to a class of algorithms related to Q-learning. We also focus on model-free algorithms, which do not attempt to construct a direct model of the environment that can be queried, but rather to derive an optimal response for a given situation, referred to as a policy. This seems especially apt as it tests the general abilities of RL algorithms and still allows to incorporate model-based domain knowledge in future research. Q-learning algorithms are part of value-based algorithms that derive the policy by taking actions that are expected to yield the highest future reward and the learning task is focused on optimising the estimates of these values. This stands in contrast to policy-based algorithms, which aim to optimise the policy directly. Although, Q-learning algorithms have a more limited field of application than some policy-based methods, they still allow for rich insights into the possibilities of RL algorithms in economic applications. 

\subsection{Stochastic Games}

To formalise games, we rely on the standard notations from finite Markov decision (MDP) processes. Players in a game are referred to as agents, indexed by $i=1,\dots,K$ and in this context and the environment is the set of factors outside of the agents direct control. For the purpose games, the finite set of states $\mathcal{S}$, is constructed from the actions of all players that are aggregated at every time step  $t=1,2,\dots$. For notational simplicity, we assume that each agent $i$ can chose from the same action set $\mathcal{A}_i$ at every time step. Hence the joint actions $\boldsymbol{a}=(a_1,\dots, a_k)\in\mathcal{A}=\mathcal{A}_1\times\cdots\times\mathcal{A}_K$ also defines the state set. Transitions from one state to the next are regulated by a state transition model $\mathbb{P}[s_{t+1}|s_{t}, \boldsymbol{a}_t]$. Each agent also possesses a reward function $R_i: \mathcal{S}\times\mathcal{A}\rightarrow \mathbb{R}$ which corresponds to the one-period (expected) pay-off received in a standard game. For an observed state $s_t$, each agent chooses its action according to its policy $\pi_i(s_t)$ and then receives the reward based on the current state, its action $a_i$ and the actions of all other players $\boldsymbol{a}_{-i}$, that is $R_i(s_t, a_i, \boldsymbol{a}_{-i})$. By setting a discount rate $\gamma \in [0,1)$, and given an agents policy $\pi_i$ and to policies of the other players $\boldsymbol{\pi}_{-i}$, the state value, that is the expected value of a state if all agents follow their policy, is defined as
\begin{align}\label{eq:state_value}
    v_{\pi}(s) &= \mathbb{E}_\pi\left[\sum_{t=0}^\infty \gamma^j r_{t+1+j}\big|s_t=s\right] \nonumber \\
    &= \sum_{a_i \in \mathcal{A}_i}\pi(a_i|s) \left[ r_i(s, a_i,\boldsymbol{a}_{-i}) + \gamma \sum_{s^\prime \in \mathcal{S}}\mathbb{P}[s^\prime|s,\boldsymbol{a}]v_\pi(s^\prime)\right]\enspace,
\end{align}
where the equation is referred to as the Bellman equation of the value function. Note that the policy here needs to be known, as the further dynamics are partially specified through it. Were the environment dynamics specified by the transition function and the policies of the other agents known, this problem could be solved by dynamic programming. As this is in general too complicated, RL algorithms provide a possibility to estimate these values using sampling techniques. Of particular importance for what follows is the approach where the actions taken by the other players are counted as part of the environment. As we will show, this has a large influence on how the the results can be interpreted, as it is generally assumed that the policies of the other agents are fixed. 

\subsection{Q-Learning}

Instead of modelling the value of a state, one can chose a slightly more granular metric, the state-action value. If the set of actions is finite, this allows to evaluate all possible actions at every state and hence chose the most profitable action among the set of possibilities. The state-action value for a given policy $\pi$ and a given action $a$ can be expressed as
\begin{align}
    q_\pi(s,a) = \mathbb{E}_\pi\left[\sum_{j=0}^\infty \gamma^j r_{t+1+j}\big|s_t=s, a_t=a\right]\enspace. \nonumber 
\end{align}
Note that here following the policy in future steps is important to calculate the given values, but it allows to evaluate every action at a given time step. Whereas the optimal policy could be derived in the case where all parameters are known, RL is generally used in applications where the environment is too difficult to understand. This leaves the two possibilities to either model the environment (model-based) or sample from the environment and bypass the modelling step (model free). For model free methods, the environment is sampled and each state-action value is associated with an estimate its Q-value. Q-learning is an algorithm that effectively updates the state-action value estimates, which is then used to derive an optimal policy, that is a policy that optimises the long-run reward $\sum_{t=0}^\infty\gamma^{t+1} r_t$. If the state-action values are correct simply choosing the action associated with the highest Q-value at any given state produces this optimal policy. Here, this policy is referred to as the greedy policy. However, if the values are incorrect, which happens for example at the start of learning, an agent needs to explore the action space sufficiently. A possibility would be to sample the environment by choosing random actions and then estimate the associated state-action values. For games this poses two problems however. First, as dynamic games rely on the reactions of other players, assuming all players chose random actions all the time effectively prohibits the learning of longer term common strategies, which does not seem realistic. Further, by randomly sampling actions, each agent potentially forgoes many rewards by choosing sub-optimal actions. This is related to the exploration-exploitation trade-off, where sufficient exploration is needed in the training phase to ensure correct estimates, and exploitation yields the (currently estimated) highest rewards. A simple strategy often used in RL is to employ an $\varepsilon$ greedy strategy. That is, with a probability of $\varepsilon$ a random action is sampled to explore the space of possible state-actions and with a probability of $1-\varepsilon$ the greedy action is chosen to minimise foregone rewards. Typically, at the beginning of the learning process, an agent should focus more on exploration and then minimise exploration at later stages of the training process to focus on exploiting its learned estimates. This can easily be achieved by setting $\varepsilon$ as a function of the training steps, that is
\begin{align*}
    \varepsilon_t = \exp(-\beta t)\enspace,
\end{align*}
with $\beta$ a hyperparameter and $t$ the training step (here, the round of a game being played). Q-learning, developed by \cite{watkins1992q} relies on a strategy that also allows to minimise the required sampling steps by employing one of the fundamental achievements within the field of RL, temporal difference (TD) learning. Briefly summarised TD, combines the idea of dynamic programming, that is, using estimates from other states to calculate the value of the current states and sampling. Whereas standard sampling techniques based on bootstrapping work well in many static problems, RL problems make this exceedingly difficult due to the environment dynamics. Here, a transition from one state to another is merely a part of a sequence, and hence sampling would need to be extensive to cover all possible sequences for a given state set. In its simplest form, TD exploits previous estimates by correcting the estimate from the current state but treating the value estimate of the next state as given. Formally, given a state $s_t$, a reward $r_{t+1}$ and the next state $s_{t+1}$ the TD error is defined as
\begin{align*}
    \delta_t = r_{t+1} + \gamma v(s_{t+1}) - v(s_{t})\enspace,
\end{align*}
which can be thought of as the error made in the estimation of the current state that is known once the immediate reward has materialised. This error can then also be used to update the estimates of the Q-values, which results in the Q-learning algorithm:
\begin{equation}\label{eq:qlearning}
    q(s,a)^{(t+1)} = q(s,a)^{(t)} + \alpha\left[r_{t+1} + \gamma \underset{a}{\text{max}} q(s_{t+1}, a)^{(t)} - q(s,a)^{(t)}\right]  = q(s,a)^{(t)} + \alpha \delta_t \enspace,
\end{equation}
where $\alpha$ is a hyperparameter referred to as the learning rate and the policy $\pi$ is included implicitly. For example, the current policy can be $\varepsilon$-greedy, but the key change in the updates is that the current Q-value is updated under the assumption that for all future periods the agent would follow the greedy policy. Although the algorithm is simple in nature, it provides a powerful tool for model-free reinforcement learning and many future refinements that are discussed below rely on this basic strategy. 

\subsection{Function Approximation}

The seminal work of \cite{calvano2020artificial} relies on the simple tabular method of Q-learning as describe above, where all possible Q-values are collected in a Q-table, which represents the knowledge of the agent about its environment, which is referred to as the tabular approach. In this section we will expand upon this with slightly more complex methods. We aim to provide a short introduction to the conceptual aspects used in RL, rather than aiming to provide a complete overview of relevant methods. For a more complete overview of the subtleties, interested readers can find additional information in \cite{rl_book} and the articles referenced. 

The tabular approach has two key weaknesses. First, tabular methods require the agent to hold the entire knowledge in memory in order to derive an optimal action. For example in the setup of \cite{calvano2020artificial}, states are defined through the actions of all players in the past iteration. Even in the simple setup, where each agent only has 15 possible actions, the Q-table for a two-player game is already of size 3375 ($|\mathcal|\cdot|\mathcal{S}|$ and $|\mathcal{S}|=|\mathcal{A}|^{K}$). Further, although it is sufficient the consider only the last state, oftentimes the agents are able to learn faster by including a larger state memory $L$, that includes past states up to time $t-L$. This would make the state space even larger as $|\mathcal{S}|=|\mathcal{A}|^{K\cdot L}$. This leads to an enormous increase in state-action space, making it difficult for the agent to learn anything without many exploration episodes. A second weakness is that the tabular nature ignores similarities across states. For example, if one state is only superficially different from another, tabular approaches cannot learn from the other states as each entry in the Q-table is isolated. Function approximation can help overcome both of these issues. The key idea is that instead of learning each state-action value individually, an agent can compress the knowledge using an approximate value. Here instead of the state itself, features are derived from each state and passed to a prediction function. Formally, for some set of parameters $\theta$, the goal is to find a good approximator $\hat{q}(s,a;\theta) \approx q_\pi(s,a)$. A popular approach is to replace the Q-table with a neural network, which was used by \cite{mnih2015human} to achieve a well performing agent in a complex setting but the choice of model can also be a simpler linear model. 

\subsection{Learning in Economic Games}\label{subsec:econgames}

In economic games, learning can be thought of as deriving an optimal action based on the current state. Similar to static one-period games, this involves taking into consideration pay-offs from each action. However, due to the dynamics of the problem, this also involves all future pay-offs, which allows to incorporate strategies of the other players (additionally to the difficulty that future rewards need to be estimated). A simple and well-known problem that illustrates why cooperation might be difficult to achieve in non-repeated games is the \emph{Prisoners dilemma}. The basic setup involves two players that can either choose to cooperate or to defect from a strategy that would optimise the total payoffs from the game, as illustrated in Table \ref{tab:rl:payoff_single}. The problem is that this type of game possesses a single Nash equilibrium\footnote{Informally, a Nash equilibrium is achieved when every player in a game does not have an individual incentive to change their strategy.} where both players defect in a non-repeated game. If the game is repeated, a \emph{tit-for-tat} strategy, is shown to be a stable solution instead \cite{axelrod1981evolution}, here the optimal strategy is to cooperate as long as the other player does so as well.

\begin{figure}[htbp]
\centering
\includegraphics[width=\textwidth]{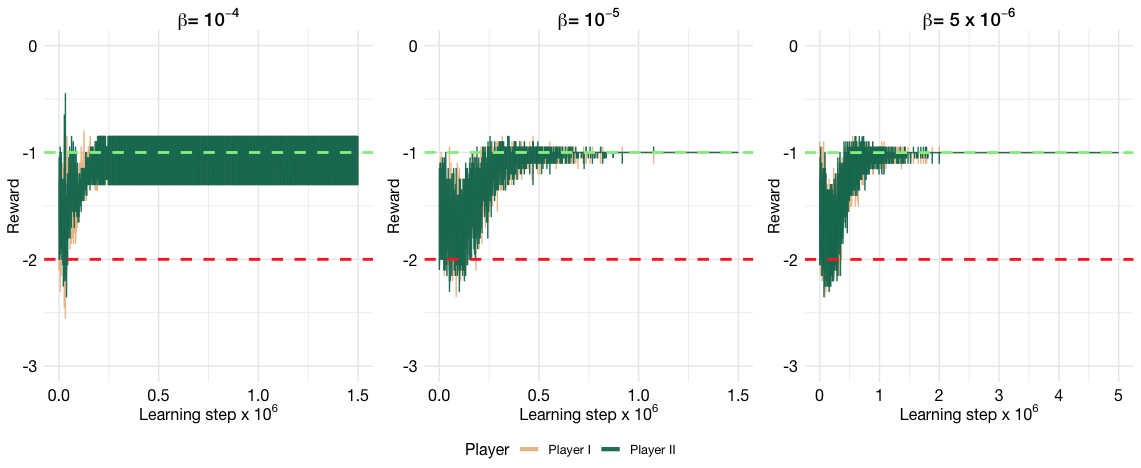}
\caption[Iterated Prisoners Dilemma in RL]{Iterated Prisoners Dilemma using two RL agents with different exploration decay rates. Note that in the right plot the $x$-axis is longer than in the other two, as the agents needed more time to converge. In red is the expected payoff under the Nash equilibrium and in light green the expected payoff under the Pareto- optimal outcome.}\label{fig:game:prisoners}
\end{figure}

\begin{table}[]
\begin{center}
    \def\arraystretch{1.25}%
    \begin{tabular}{ccc}
     \multicolumn{3}{c}{Payoff Matrix}  \\
    \hline
    \hline
     &  \multicolumn{2}{c}{Player II}  \\
     \cline{2-3}
     Player I & Cooperate & Defect  \\
     \hline
     Cooperate & $\rho_{C}, \rho_{C}$  & $\rho_L, \rho_T$ \\
     Defect & $\rho_T, \rho_L$ & $\rho_D, \rho_D$\\
     \hline
    \end{tabular}    
\end{center}
\caption[Payoff Matrix for Prisoners Dilemma]{Payoff matrix for a single prisoners dilemma, $\rho_T > \rho_C > \rho_D > \rho_L$}
\label{tab:rl:payoff_single}
\end{table}
Within the context analysed here, this can easily be translated as a reward optimising problem that is solvable by a reinforcement learning agent.
\begin{align*}
    \sum_{t=0}^T \gamma^t r_t = \sum_{t=0}^{t_0} \gamma^t \rho_C + \sum_{t=t_0}^{T} \gamma^t \rho_D  
\end{align*}
Although these results are well-known in theory, it allows us to analyse the dynamics that different components of reinforcement learning algorithms. We run a first set of experiments by fixing the parameters for $\rho_T, \rho_C, \rho_D, \rho_L$ as $\lbrace 0,-1,-2,-3 \rbrace$ and train a simple tabular Q-learning agent using repeated interactions. The discount rate $\gamma$ is set to $0.95$ and the two other parameters that need to be chosen, the learning rate $\alpha$ and the exploration rate $\varepsilon$ are varied across the experiments. As the exploration rate $\varepsilon$ needs to converge towards zero with increased training time, we follow the standard approach and set $\varepsilon = \exp(-t \beta)$, where $\beta$ is a hyperparameter and $t$ is the training step. All experiments are averaged across 20 Monte-Carlo iterations, wherein the initial values of the Q-table are chosen using random uniform sampling. The results are visualised in Figure \ref{fig:game:prisoners}. 
The results are largely as expected and visualised in Figure \ref{fig:game:prisoners}. Both agents learn to cooperate effectively. It is worth noting however, that even in such a simple game the algorithms take time to converge to a stable equilibrium. Further, the learning process is extremely expensive and although an algorithm might not converge to a unique equilibrium (such as for the case with $\beta=10^{-4}$ in the left pane), the total reward after one million iterations is still around 17\% larger than that received by an agent that converges to the optimum, but at a slower rate. Although overall, the choice of multiple hyperparamters is important, much of it comes down to the exploration rate. This problem is well-known in many applications of RL but has paramount importance in applications for economic games. Whereas in many common applications of RL, such as robotics, a longer training time means higher costs in terms of compute units, in economic games it can result in results that converge but have little economic sense. For example, in the iterated prisoners dilemma, this would mean an agent learns to cooperate only after having spent many lifetimes in prison already.

\subsection{Learning in Pricing Games}

The insights from the very simple prisoners dilemma are important for what follows, as typically economic pricing games, such as considered by \cite{calvano2020artificial,rocher2023adversarial} are more complex and difficult to resolve with other methods. Previous literature on the topic has mainly discretised the space of possible actions by imposing limits on the number of actions that can be taken. We largely follow this approach here but will sometimes the price values themselves, rather than its state representation for tasks involving function approximation. Note further, that there also exist approaches to deal with continuous problems, for example \cite{lillicrap2015continuous}, but we will abstract from this possibility here to ensure comparability with previous results.

In general, we focus on the setup where $K$ agents compete in a market environment where demand functions are unknown to the agents. The agents compete for market share and try to optimise their payoff by proposing prices $p_i$ for a product that are higher than their production cost (here, normalised to 1). The market environment, identical to the one used in \cite{calvano2020artificial}, is specified as
\begin{equation}\label{eq:demand}
    q_i = \frac{e^{\frac{a_i-p_i}{\mu}}}{\sum_{j=1}^K e^{\frac{a_j-p_j}{\mu}} + e^{\frac{a_0}{\mu}}}, \quad i=1,\dots,K,
\end{equation}
where $a_j, j=0,\dots,K$ are product specific quality indexes and $\mu$ a differentiation parameter (both fixed outside of the control of agents) and $e^{\frac{a_0}{\mu}}$ represents demand for an outside good. The goal of every firm is to adjust the price to maximise its profit, defined as $\pi_i = (p_i - {MC}_i) q_i$. From economic theory, it is well-known that competition within the market should drive prices down, as lowering high prices on a concave demand function results in significant increases in the demand. A market is said to be in equilibrium, if no producer has an incentive to change its price, which translates to having a constant action choice for all $K$ agents in the game. 

The simple economic environment further has the advantage that the theoretic monopoly price, collusion price and competitive price (the Nash equilibrium) can be calculated, the derivation is straightforward and relegated to the supplementary materials. The prices for competition and monopoly are denoted as $p_C$ and $p_M$ respectively for the parameters they use in their experiment and then linearly interpolate between these two prices. For the first set of experiments, the environment parameters are identical to \cite{calvano2020artificial}. That is $\mu=0.25, a_0=0$ and $a_i=2, MC_i=1$ for $K=2$. To discretise the action space, we first calculate $p_C$ and $p_M$ and interpolate linearly between $[p_C - 0.1(p_M-p_C), p_M + 0.1(p_M-p_C)]$, again, identical to \cite{calvano2020artificial}. To learn from the game, it is assumed that each player in the game can observe the actions of other players, but that the reward is private to a single agent. The current state can then be expressed by the past action of all players, which enables to incorporate basic strategies such as tit-for-tat in the state information. To formalise this, we consider the case where each agent in the game $i=1,\dots,K$ has a potentially unique state value function $V_i(s)$ and an associated action value function $Q_i(s,a)$, but all players find themselves in a common state defined by the action memory. Learning in this context can be considered the experience that an agent has accumulated about the actions of other players, given its own action set. By considering the actions of other players as part of the environment, each agent does not need to model the actions of the other players separately. However, not all research on the topic relies on this setup. For example, the approach taken by \cite{rocher2023adversarial} could be considered a strategy to model the actions of the other players rather than the environment as a whole, although their approach is strictly speaking not an RL problem. This will be expanded on below as well.

The $Q-$table or $Q-$matrix of each player then represents the set of state action values that are currently estimated. For simple games, this setup also allows a direct interpretation of the values within the $Q-$table. As an example, consider a simple $2\times 2$ normal form iterated prisoners dilemma with a memory of length $L=1$. Then the value of each entry in the $Q-$table represents the discounted value of each action given the current strategy. For larger problems, scenarios are usually sampled once the algorithms have converged within the game. 

\section{Pricing Decisions and Learning}\label{sec:main}

The two key questions that naturally arise when studying algorithmic pricing are the following. First, are the algorithms actually {\em colluding} or do they merely converge to a given equilibrium. Second, is their implementation realistic under regulatory constraints? The first question will be addressed by interpreting sampling results in different situations. The second question is more difficult to answer, as it will be shown, much of the previous results rely heavily on convergence and symmetry in the games. Further, economic constraints are often ignored, as algorithms are usually free to explore as much as necessary in the training stages of an analysis. This facilitates the convergence towards equilibria that resemble the situation of collusion. However, it would also result in substantial losses or lost profits throughout the training phase. In many industries, this would inevitable lead to either bankruptcy of the firm in question, or at least a change in operative leadership. Hence, if an algorithm is to be considered realistic, it needs to converge towards an equilibrium \emph{fast} and also without \emph{any agreement} on the deployment, as otherwise existing regulations prohibiting their use might apply. 

\begin{figure}[htbp]
\centering
\includegraphics[width=\textwidth]{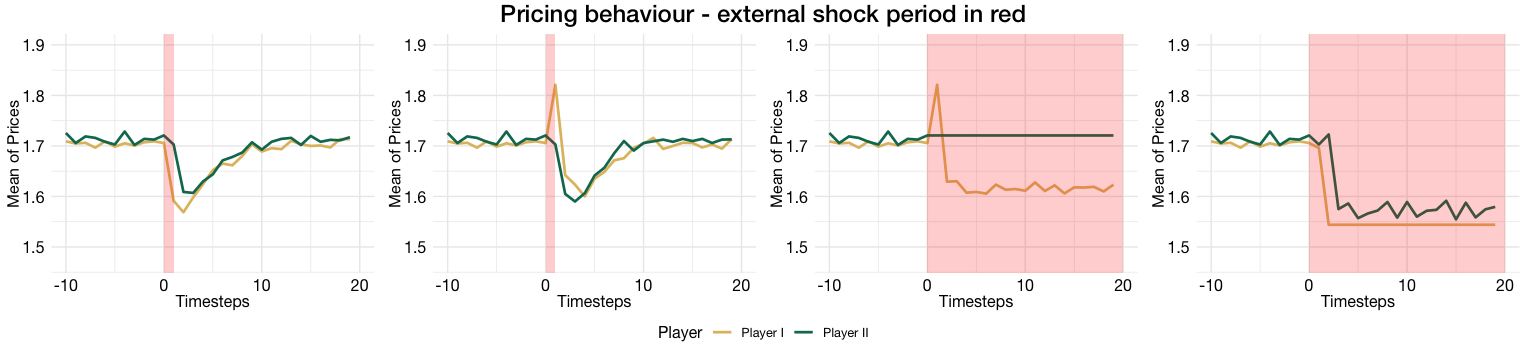}
\caption[Response Function of Converged Agents]{Response function of two agents that have converged in their training. The lines depict the mean across the trained agents, the red shaded area shows where we manually overwrote the actions chosen by the policy of at least one agent. }\label{fig:game:response}
\end{figure}

We investigate these points through a series of experiments. As a first step, we train two Q-learning agents in the same way as in the reference article of \cite{calvano2020artificial}. Instead of focusing on the hyperparameters $\alpha$ and $\beta$, that is, the setup of the agents, we focus on the interpretation of the outcomes when the setup itself is slightly changed. As argued in \cite{calvano2020artificial}, in order to detect collusive behaviour of algorithms a convergence to higher prices is not sufficient. They employ a simulation-based approach that we reuse here, but expand upon. Results are visualised in Figure \ref{fig:game:response}. The left pane shows essentially a replication of \cite{calvano2020artificial}. Here, two algorithms that have converged are exposed to a shock at time $t=0$. We manually overwrite the response of one of the agents to propose a lower price for a single period (market in red) and observe the proposed prices of both agents for the following periods. The ensure that the results are robust, we average the obtained price response across 80 Monte Carlo iterations that differ in the initialisation and the exploration during training. The results are in line with what \cite{calvano2020artificial} found, if an agent deviates the other agent responds by also reducing its price. The price then rises gradually back to the (collusive) equilibrium price. In previous studies, this is interpreted as the algorithms having a punishment phase, where the second player lowers its price to limit the profits of the other agents. The centre panes depicts a slightly different result when we manually force a higher price on one of the agents. Here the results are somewhat surprising, as the second agents again lowers its price, although one would expect it keep it steady instead (as it would result in a higher market share, at the same price). The centre right pane depicts the reactions when the first agent temporarily increases its price, but the second agent is forced to hold the price. This results in the first agent lowering its price below the pre-shock levels, which would indicate that the strategies are more dependent on changes of the other agents behaviour, rather than an actual understanding of the market dynamics. The right pane shows how the agents react if we permanently set the price of one of the agents to just above the Nash price. Here, again the results are somewhat surprising, as the second agent first drops its price, but not enough counteract the effects of the lower price of the competition, in turn losing most of the market share.
The observed behaviour might put into question some results from the literature, as it appears that the algorithms simply learn to converge back to the equilibrium price after a simple shock rather than having a truly collusive strategy. Further, the learned strategy appears to imply a limited understanding of the mechanisms, where lowering prices is understood, but not raising prices. A possible remedy would be to have more complex models, which would probably result in longer training times, but this would expose the algorithms to the critique of not being realistic enough. In turn, it appears that the algorithms are easy to fool and would present a significant risk for any firm employing them without close supervision. 

\subsection{Non-Symmetric Hyperparameters}

A further key assumption that was made throughout previous studies but rarely discussed is the symmetry of hyperparameters of the agents. This is generally justified by assuming that every firm has access to the same technology. Although this makes simulating results particularly convenient, it comes with multiple issues. First, as one key hyperparameter is the discount rate, it would be assumed that every agent has exactly the same preferences with respect to the future. This seems irrealistic, especially in the case where two competitors do not have the same maturity in the market. Second, exploration is crucial for convergence to a collusive equilibrium as shown above. In general, every possible state and every possible sequence of states should be visited often enough, such that an optimal policy can be derived by the agents (which would mean having a collusive strategy in this setup). This was formalised in recent research by \cite{banchio2022adaptive}, who found that collusive equilibira are essentially the result \emph{spontaneous coupling} in the training process, that help discover the long-run profitable equilibrium. Here, the exploration-exploitation trade-off can be understood in a simple manner. Where too few visits to short-term sub-optimal states result in a long-term sub-optimal outcome. This has important implications as in economic environments having a try-out phase where algorithms can freely explore, seems unrealistic. This would essentially mean that a company would need to forgo profits, which seems difficult to justify before either a board or shareholders. By setting the exploration rate symmetrically for all agents, at every timestep this adds an additional constraint on the time of the deployment of an algorithm. \cite{rocher2023adversarial} address this issue by constructing a graph-based approach that focuses on finding profit-maximising states from a stationary competitor with few iterations. This allows a more natural approach, where instead of being trained together, it is assumed that the algorithms are already deployed. However, as will be discusses below, it relies on the discretized nature and stationarity of the policy of the opponent. 

\begin{figure}[htbp]
\centering
\includegraphics[width=0.8\textwidth]{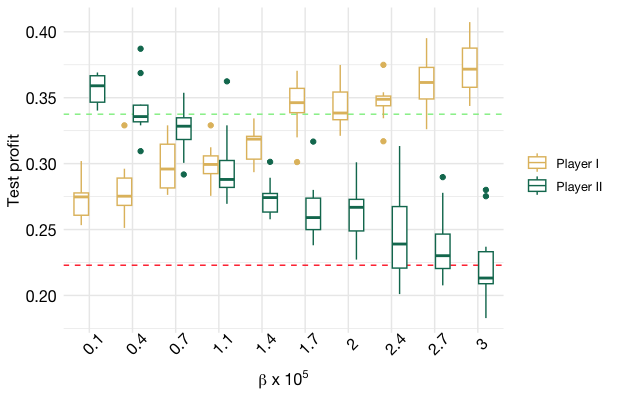}
\caption[Results for Non-Symmetric Hyperparameters]{Rewards after training with different exploration rates. Player I has a fixed exploration rate set as $\varepsilon=\exp(-t 10^{-5})$ and we vary the exploration rate of the second player. The dotted lines in green and red represent the symmetric collusion respective Nash profits.}\label{fig:game:nonsymmetric_hyper}
\end{figure}

To investigate the influence of this assumed symmetry, we conduct a short experiment. We fix the exploration rate of one agent and vary the rate of the other player. This allows us to analyse the profits obtained for each player across the situations. Intuitively, both agents initially explore the space but one of the agents has a faster exploration decay rate than the other. This could be interpreted as one agent being more confident of having explored the space enough. A second way to interpret this outcome, has a more practical foundation, as it can be considered a game in which one player can update its strategy faster than the other. For example, a larger player with a better pricing infrastructure can update its policy twice as fast. This gives a realistic interpretation to the results, as in real markets there is usually a player with more resources. The results of this investigation are summarised in Figure \ref{fig:game:nonsymmetric_hyper}.
The results are fairly consistent across the different changes. It shows that having a slower learning agent, that is a lower exploration rate decay, is beneficial for the long term profit. A simple way to look at this is that an agent which can still learn once the other agent has converged, can simply exploit the static behaviour of the other. As compared to previous studies, this indicates that algorithmic collusion has a large dependence on the symmetry of the hyperparameters. In general, there is no reason why this should be the case in general applications. A possible argument is that several companies could be using the same pricing algorithm, however, our results cast doubt on the feasibility having the \emph{exact} same deployment without any prior agreement. However, such an agreement would result in a direct collusive behaviour as mentioned in the introduction above. Further, it appears that such algorithms are not robust to non-stationary pricing algorithms, which might make them susceptible to being exploited by an opponent. 

\subsubsection{Opponent Modelling}

Using the previous findings, we can state two key observations. First, achieving a collusive behaviour with simple RL agents relies heavily on the symmetry of the parameters of the agents and second, the results rely on agents being either stationary or reaching stationarity at the same time. This makes previous studies subject to criticism, as firms under the given economic constraints would find it hard to implement such algorithms. However, the conclusions drawn, mainly that this overwhelmingly results in bad market outcomes for consumers can still remain valid. Here, we expand upon previous research and devise a more general framework in which collusion or collusive outcomes can be reached. Essentially, we will follow a similar idea to what \cite{rocher2023adversarial} proposed and consider agents that can observe the behaviour of other participants in the market, but propose an approach that is also valid for more general settings. In what follows, we refer to an agent that can be observed in the market as the incumbent and the agent we train on the observed behaviour as the newcomer. Relying on market symmetry usually requires every participant in the market to have the same pricing algorithm. To depart from this assumption and allow more complex policies for the incumbent, we rely on function approximation instead the tabular representation. This also allows us to incorporate more information in the state representation. As a second step, we re-address the issues arising due to the constraints imposed by profit maximisation, namely the long training times and possibility to lose money if a competitor changes its pricing model. We do this by investigating how an agent can more effectively balance the exploration-exploitation trade-off and learn from incumbent firms that might change their pricing policy in the future. 

\begin{figure}[htbp]
\centering
\includegraphics[width=0.8\textwidth]{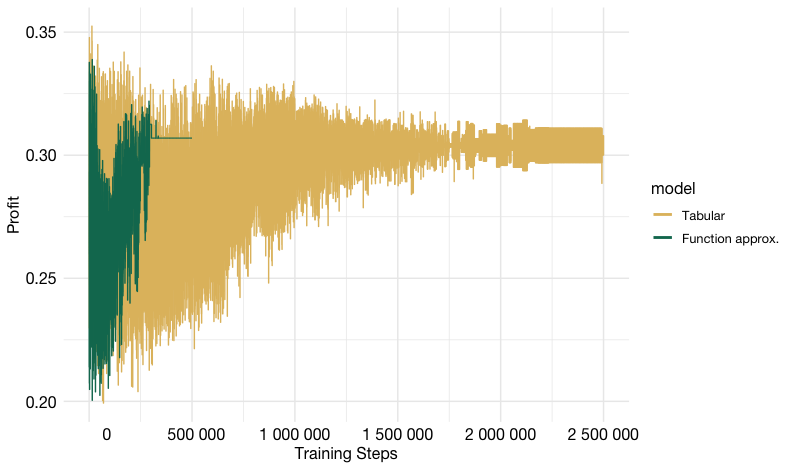}
\caption[Agent Training in General Setting]{Convergence speed in a symmetric market with the tabular approach and a function approximation approach until convergence is reached. The function approximator takes into account the last ten states which appears to increase the convergence speed. }\label{fig:game:extensions}
\end{figure}

To develop a more general agent using function approximation, we construct our agent using a Deep Q-Learning approach developed by \cite{dqn_mnih}. Here, the Q-table is replace by a neural network that approximates the state-action value in much the same way as the tabular method. However, as it allows to incorporate a larger information set, we can increase the state memory length $L$. This in turn should help agents converge faster, as it is easier to differentiate between different trajectories, making identifying the policy of the other agent easier. First, we consider the simple setting where an agent can observe the last ten states from the incumbent and adapt its policy accordingly. The incumbent is then set up in much the same fashion as the newcomer and we analyse the convergence speed, which we hope to increase. Indeed, as depicted in Figure \ref{fig:game:extensions}, in the standard setup, the agents converge faster on a collusive equilibrium, especially if the state-memory is increased. This would not be feasible with a tabular representation, as even if the price space is discretised to only 15 possible values, a two-player game with a 10-state memory would already possess over $10^{23}$ possible states. This further motivates us to devise a more general agent using function approximation instead of a tabular representation. Function approximation also enables us to have a more clear-cut interpretation of the underlying mechanism. Although the state space is discretised, it is not an assumption of the DQN agent, and hence we can conduct sampling with smaller differences to understand the inner workings of the neural network better. 

However, as discussed above, even such an agent is sensitive to the stationarity of the environment as even a more complex agent needs to explore possible options. Further, although the DQN agent seems to converge faster it still takes many iterations until a stable price is reached. Similar to what happened in the prisoners dilemma above, this might be infeasible due to constraints imposed on the economic viability. It also misses an important point, namely that in reality, a market does not have to start with zero observations. To address this issue we rely on a novel algorithm that uses both online (sampled) information and offline information (that is, past observations), inspired by recent research~\cite{zheng2023adaptive}. The goal here is to create an agent that can effectively balance the two types of data to avoid long training times but also to adapt its policy in case the incumbent changes its pricing strategy, which we saw above leads to an irrational behaviour in tabular agents. The key idea, inspired by \cite{zheng2023adaptive}, is to have two types of updates to the value function. Updates are performed based on a buffer of past observed data, that is given less weight, and observations from an online-buffer that are sampled on the go and assigned more weight. As compared to the original paper by \cite{zheng2023adaptive}, we replace the probability to sample from either buffer from a fixed sampling rate, to one that depends on the difference in sampled observations.

In more detail, we consider the situation where an agent (again, referred to as the newcomer) updates its pricing model from a tabular, pre-trained model similar to the setup of \cite{rocher2023adversarial}. Instead of deriving a specialised method to model the behaviour of the incumbent, we instead rely on a DQN agent with more general learning capabilities that does not rely directly on the setup of the environment. Here, a newcomer first observes the market in equilibrium and fills its offline data buffer of size 4000 with the observations. It then changes its base model and starts training, in turn filling the online-buffer which has a size of 400 to keep information current. The probability to sample from either buffer is then dependent on a rolling profit average metric, which causes the probability of sampling from the online buffer to increase whenever it falls below a threshold, which in turn automates a human supervision. 

\begin{figure}[htbp]
\centering
\includegraphics[width=\textwidth]{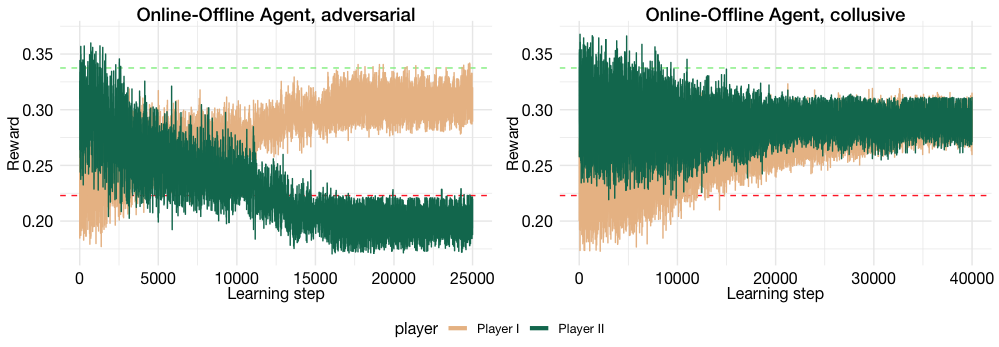}
\caption[Online- Offline Agent]{Market outcomes when an online- offline agent is used under adversarial (competitive) and collusive scenario}\label{fig:game:onlineoffline_stationary}
\end{figure}
\begin{figure}[htbp]
\centering
\includegraphics[width=\textwidth]{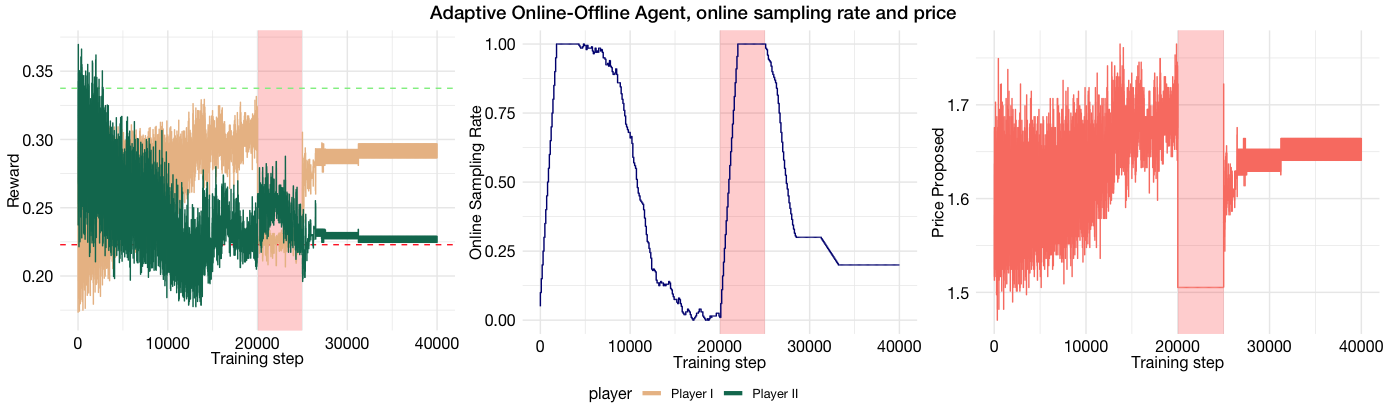}
\caption[Online- Offline Agent with Adaptive Sampling Rates]{Market outcomes when an online- offline agent is used under the adversarial scenario if the online sampling probability is adaptable. The red zone marks where the incumbent agent drops its price to just above the Nash price.}\label{fig:game:onlineoffline_nonstationary}
\end{figure}

At first, we study how quickly the new agent can exploit the market when its opponent keeps a stationary policy. The results are visualized in Figure \ref{fig:game:onlineoffline_stationary} with Player I being the newcomer. Similar to \cite{rocher2023adversarial}, we consider both the adversarial situation, where the agent tries to exploit the stationarity of the other algorithm and the collusive situation, where the agent tries to maximize its profit as well as the opponents profit. This can be achieved by giving the market share of the opponent a weight in the reward function which effectively combines multiple objectives into a single optimisation problem. The convergence towards both an adversarial and collusive equilibrium is extremely fast, although the general approach takes more training steps than the more task specific solution of \cite{rocher2023adversarial}. However, as discussed above, these algorithms largely rely on the stationarity of the policy of the opponent. To put this to the test, we extend the example. Now, the incumbent will change its policy temporarily to a price just above the Nash equilibrium price, which was shown above to confuse simple agents. The results of this experiment are summarised in Figure \ref{fig:game:onlineoffline_nonstationary}. As can be seen, the new agent can effectively deal with the nonstationarity in the incumbent agents policy and immediately and stably drops its price. It re-adapts its sampling to give more weight to recent observations but keeps on learning until it eventually reaches a new equilibirum, which is again changed once the incumbent readapts its policy. This in turn results in both higher short-term and long-term profits for the agent using a more flexible approach. This in turn makes reaching supra-competitive prices more likely and also more robust, validating and strengthening much of the insights of previous research.

\section{Discussion}\label{sec:conclusion}

Algorithmic pricing, and especially tacit collusion via pricing algorithms has become a popular topic of research. Reinforcement learning provides a powerful tool to study situations that were hereto too complex to be modelled with a general class of models. However, given that derived policies are difficult to interpret, studies have to rely on an interpretation of the results via sampling, which makes many insights subject to criticism of either incomplete study setups or overdrawn conclusions. In this article, we discussed RL in a more general context of games and showed how simple games can be extended to more complex ones used in the literature. We found that, although trailblazing in the general nature of the problem, previous research heavily relied on two key assumptions, namely that policies are stable and that there is an irrealistic amount of symmetry necessary to achieve results that lead to outcomes akin to collusion. 

However, the general insight, namely that market outcomes appear to drive away from results deemed to be generally in line with older competition theory, seems to hold true with a few tweaks to the algorithms. Whereas this might be referred to as true collusive behaviours of RL-agents remains to be discussed, but our findings indicate that the underlying mechanism should be considered under the angle of an optimisation problem. Cartels were (or are) difficult to implement due to the sluggishness of the reaction times of co-conspirators, which makes deviating particularly beneficial. Further, as discussed in the introduction, setting up a joint structure in the very beginning of cartel formation poses a significant obstacle for would-be cartel members. Both of these issues can be effectively addressed using modern RL algorithms. In RL-driven games, powerful agents can react momentarily to changing conditions and make deviating of other agents easily accessible. Even if punishing strategies are difficult to obtain out of the box, algorithms can easily integrate such strategies with a few simple tweaks. Using powerful base-learners also cuts down on the required training time necessary to achieve a convergence on values, which lowers the entry costs. The general insights have lead to many policy recommendations, of particular importance is the discussion in \cite{clark2023algorithmic} who suggest that market authorities could change their approach and advises that ``{\em rather than pursuing hard-core cartels on an individual basis, it might be more effective to concentrate on collusion-facilitating devices that do not even require a conspiracy, such as AP} [Algorithmic Pricing] {\em and communication via earnings call}''. Our research here contributes to exactly this undertaking. As a blanket ban on pricing algorithms seems unlikely, more research is needed on the possible applications where algorithmic pricing leads to anti-competitive behaviour between firms. Further, our findings expand on what \cite{rocher2023adversarial} argued in their research. Instead of having truly collusive strategies, it appears that algorithms simply interpret signs in the market and optimise a long term goal. If this long-term goal involves the overall profit of producers, explicit collusion is not even necessary.

Although the discussion in the field currently revolves around the possibility of collusion, our simulations above show that the problem is more complex than that. With a more generally capable agent, the main driver behind the outcome is the reward function. Although the results are similar to collusive pricing strategies, it does not means that there is necessarily an intent behind the actions, nor that the algorithms used truly understand collusion and punishment. Further, it is difficult to find the difference between prices in a market that are set while observing the prices of competitors, and having as explicit goal to optimise the return by appeasing competition in the long run. This poses an issue for current policy surrounding price fixing, as it takes away the intent to keep prices artificially higher and replaces it with machine rationality. This in turn makes collusive behaviour outcomes more likely, as they emphasise the long-run return, just as many cartels would. If competitors start to adopt such algorithms to maximise long-run profits, this might result in higher prices overall. Whereas training algorithms in isolation takes a prohibitively long time to converge, innovation in agent design might cut this training time down further, in turn posing less issues with the financial constraint. Whereas previous studies often relied on hundreds of thousands of training steps, our approach here converges quickly and due to the adaptive exploration strategy remains capable of learning in a nonstationary environment.

\pagebreak

\bibliographystyle{apalike}
\bibliography{bibliography}  

\clearpage

\appendix

\section{Derivation of Equilibrium Prices}

For the given parametrisation, where all agents have the same coefficient of differentiation $a_i$, the market will be in equilibrium if $p_1=p_2=\cdots=p_K$. The upper solution (the monopoly prices) and the lower solution (the Bertrand competition case) can then be found by maximising the profit with respect to the market demand. The difference is that in the monopoly case, the market demand is only depending on the price of the supplier, whereas in the competition case the demand depends on the action of others. Given that the differentiation parameter is exactly the same in the experiments, the market will be in equilibrium if the prices are the same across the suppliers. 

\subsection{Monopoly prices}

For ease of notation, assume that $p_0=2$. The goal for every firm is to maximise the profit, which is defined as $\pi_1 = q_1*(p_1-1)$. Plugging in the demand function \ref{eq:demand}, we can optimize the value for $\pi_1$ with respect to $p_1$ (which is the choice of the supplier. This is done by deriving with respect to $p_1$ and finding the extremal point, which can be done via:
\begin{align*}
    \frac{\partial \pi_1}{\partial p_1}=-\dfrac{2\mathrm{e}^8\cdot\left(\left(4p_1-5\right)\mathrm{e}^{4p_1}-2\mathrm{e}^8\right)}{\left(\mathrm{e}^{4p_1}+2\mathrm{e}^8\right)^2}\overset{!}{=}0 \\
    \implies p_1 \approx 1.924981
\end{align*}
Note that this implies that there are essentially two monopoly players that collude perfectly. 

\subsection{Duopoly prices}

For the duopoly prices, we can assume that every agent fixes the price itself, this results in:

\begin{align*}
    \pi_1 = (p_1 - 1)q_1 = (p_1-q)\frac{e^{\frac{2-p_1}{0.25}}}{\sum_{i=0}^2e^{\frac{2-p_1}{0.25}}}
\end{align*}
Optimizing the price can again be done by imposing the first order condition:
\begin{align*}
    \frac{\partial}{\partial p_1} (p_1-q)\frac{e^{\frac{2-p_1}{0.25}}}{\sum_{i=0}^2e^{\frac{2-p_1}{0.25}}} \overset{!}{=} 0 \\
    \implies p_1 = \frac{5(e^{4(p_1+p_2)} + e^{4(p_1 + 2)}) + e^{4(p_2 + 2)}}{4(e^{4(p_1+p_2)} + e^{4(p_1+2)})}
\end{align*}
This would need further simplification as $p_1$ is still present on both sides, but by imposing the fixed point condition as well as $p_1=p_2$ which turns out to be approximately $p_1 \approx 1.472927$. This corresponds to about the 47\% increase in the price-marginal cost ratio that was mentioned in the paper. The same logic applies for the five-agent game, resulting in $p_1 \approx 1.311521$. 

\end{document}